\documentclass[twocolumn,prl]{revtex4}

\ifx\pdfoutput\undefined
  \usepackage[dvips]{graphicx}
\else
  \usepackage[pdftex]{graphicx}
\fi

\usepackage{dcolumn}
\usepackage{amsmath}
\usepackage{latexsym}

\begin{document}
\title[Short Title]{Cooper pair tunneling in circuits with substantial
dissipation:\\ the three-junction R-pump for single Cooper pairs}

\author{A. B. Zorin}
\author{S. A. Bogoslovsky}
 \thanks{Permanent address:
Laboratory of Cryoelectronics, Moscow State University, 117899
Moscow, Russia.}
\author{S. V. Lotkhov}
\author{J. Niemeyer}

\affiliation{Physikalisch-Technische Bundesanstalt, Bundesallee
100, 38116 Braunschweig, Germany}%


\begin{abstract}
We propose a circuit (we call it R-pump) comprising a linear
array of three small-capacitance superconducting tunnel junctions
with miniature resistors ($R > R_Q \equiv h/4e^2 \cong
6.5$~k$\Omega$) attached to the ends of this array. Owing to the
Coulomb blockade effect and the effect of dissipative environment
on the supercurrent, this circuit enables the gate-controlled
transfer of individual Cooper pairs. The first experiment on
operating the R-pump is described.
\end{abstract}

\maketitle

\section{Introduction}
The Coulomb blockade effect in circuits with small-capacitance
tunnel junctions provides the means of manipulating single charge
quanta (see, for example, the review paper by Averin and Likharev
\cite{Averin91}). If periodic signals of frequency $f$ are applied
to the gate electrodes of the circuit, a train of single charges
$q$ can move across an array of junctions so that charge pumping,
giving rise to the current $I=qf$, is achieved. It was
experimentally proven that the normal-state metallic circuits
enable single electrons ($q=e$) to be effectively pumped at
frequencies $f$ of about several MHz \cite{Pothier}. Moreover,
the accuracy of single-electron pumping can nowadays meet the
requirements of fundamental metrology, $viz.$ $\delta I/I \simeq
10^{-8}$ \cite{Keller}.

Unlike pumping of electrons, the pumping of Cooper pairs ($q=2e$)
in superconducting circuits has not been that successful so far.
The only experiment had been carried out in 1991 by Geerligs {\it
et al.} \cite{Geerligs} with a three-junction Al sample. Although
this experiment did evidence a pumping of the pairs, the pumping
was strongly disturbed by several factors: the Landau-Zener
transitions, Cooper pair co-tunneling, quasiparticle tunneling,
etc. As a result, the shape of the current plateau at $I=2ef$ in
the I-V curve was far from being perfect.

Recently, in their theoretical paper Pekola {\it et al.}
\cite{Pekola} concluded that pumping of Cooper pairs with
reasonable accuracy in a three-junction array was impossible. For
practical values of parameters (the ratio of the Josephson
coupling energy to the charging energy $\lambda \equiv E_J/E_c =
0.01-0.1$) they evaluated the inaccuracy of pumping $\delta I/I$
to be as much as 9\%-63\%. This is because of intensive
co-tunneling of pairs in the short arrays, i.e. the process of
tunneling simultaneously across several ($\geq 2$) junctions. To
suppress the co-tunneling and, in doing so, to improve the
characteristics of the pump they proposed to considerably
increase the number of junctions $N (\gg 3$) and, hence, of the
gates ($N-1$). However, such modification would make operation of
the circuit more complex.

\begin{figure}[h]
\begin{center}
\includegraphics[width = 0.98\columnwidth]{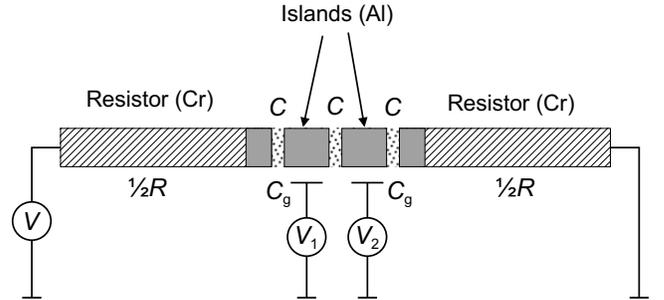}
\caption{Schematic of the R-pump for single Cooper pairs. Three
superconducting junctions of type Al/AlO$_x$/Al form two small
islands. Miniature normal metal (Cr) resistors accomplish the
structure. The device is driven by two harmonic voltages $V_{1,2}
= V_{01,02} + A \cos(2\pi f t\pm \frac{1}{2}\theta)$, where the
dc offset $V_{01,02}$ determines the centre of the cycling
trajectory in the $V_1-V_2$ plane; the phase shift $\theta \geq
\frac{\pi}{2}$.} \label{schema}
\end{center}
\end{figure}

In this paper we propose an alternative way to improve Cooper
pair pumping in a three-junction array. We modify the bare circuit
by attaching to the ends of the array the miniature resistors
(see the electric diagram of this device, which we call R-pump,
in Fig.~\ref{schema}). Their total resistance $R$ exceeds the
resistance quantum $R_Q \equiv h/4e^2 \cong 6.5$~k$\Omega$ so
that the dimensionless parameter
\begin{equation}
z \equiv \frac{R}{R_Q} \gg 1.
\end{equation}
The self-capacitance of these resistors should not be much larger
than the junction capacitance $C$. In this paper we will analyze
how such resistors affect tunneling and co-tunneling of Cooper
pairs in the three-junction array and present preliminary
experimental data.

\section{Peculiarities in operating the R-pump}
In general terms, the principle of operation of the R-pump remains
the same as that of the pump without resistors (for details of the
Cooper pair pump operation see Ref. \cite{Geerligs}). Two periodic
signals $V_1(t)$ and $V_2(t)$ are applied to the gates to form an
elliptic trajectory in the parameter plane $V_1-V_2$. At zero
voltage $V$ across the pump, this trajectory encircles in the
clockwise (counter-

\noindent clockwise) direction the triple point of the boundaries
between the stability domains for adjacent charge configurations
with an excess Cooper pair. A complete cycle results in the
charge $q=2e$ $(q=-2e)$ being sequentially transferred across all
three junctions of the array.

The Cooper pair transitions in a circuit without dissipation
(quasiparticle tunneling is neglected) occur due to Josephson
tunneling (supercurrent). The strength of the Josephson coupling
$E_J \equiv \frac{\hbar}{2e}I_c$ (here $I_c$ is the critical
current) is assumed to be $ \ll E_c \equiv e^2/2C $, the charging
energy, i.e. parameter $\lambda \ll 1$. When the domain boundary
is crossed, the resonance condition is obeyed and Josephson
coupling mixes up the two states, each corresponding to the
position of the pair on either side of the junction. (Note that,
at the same point, the resonance condition for co-tunneling in
the opposite direction across two other junctions is met.) Slow
passage across the boundary results in transition of the pair in
the desired direction. This process is mapped on the adiabatic
level-crossing dynamics (see, for example, Ref. \cite{Ao}), and
the probability of missing the transition is given by the
Landau-Zener expression:
\begin{equation}
p_{\rm LZ} = \exp \left(-{\frac{\pi E_J^2}{2\hbar \dot{E}}}
\right),
\end{equation}
where $\dot{E} \propto f$ is the velocity of variation of the
energy difference $E$ between the two states.

In contrast to the elastic (and, hence, reversible) tunneling of
pairs in the pump without resistors, tunneling in the R-pump is
accompanied by dissipation of energy. The strength of the
dissipation effect depends on the active part Re$Z(\omega)$ of an
electromagnetic impedance seen by the tunneling pair, namely by
the dimensionless parameter $z' = $ Re$Z(0)/R_Q$. If the condition
of weak Josephson coupling $E_J \ll E_c/\sqrt{z'}$ is satisfied,
i.e. $\lambda \sqrt{z'}\ll 1 $, the rate of Cooper pair tunneling
is expressed as \cite{Averin89}
\begin{equation}
\Gamma(E) = \frac{\pi}{2\hbar}E_J^2 P(E),
\end{equation}
where $E$ is the energy gain associated with this transition and
$P(E)$ is the function describing the property of the environment
to absorb the energy released in a tunneling event. (The
discussion of properties as well as numerous examples of $P(E)$
for different types of electromagnetic environment are given in
the review paper by Ingold and Nazarov \cite{Ingold}.) The
probability of a missed transition is then given by the
expression:
\begin{equation}
p_{\rm m} = \exp \left\{ -\int_{-\infty}^\infty \Gamma[E(t)] dt
\right\} = \exp \left(-{\frac{\pi E_J^2}{2\hbar\dot{E}}} \right)
= p_{\rm LZ}.
\end{equation}
Here we used the normalization condition $\int_{-\infty}^\infty
P(E)dE=1$, and the assumption of a value $\dot{E}$ constant in
time was exploited. Thus, in the case of weak coupling,
dissipation does not affect the total probability of tunneling
$p=1-p_{\rm m}$. This conclusion is in accordance with the
general result obtained by Ao and Rammer \cite{Ao} for quantum
dynamics of a two-level system in the presence of substantial
dissipation.

Using the expansion of $P$ in the region of small $E \ll E_c$
\cite{Averin89}, \cite{Ingold} one arrives at the expression
\begin{equation} \Gamma(E) \propto
\lambda^2 E^{2z'-1}.
\end{equation}
For $z' < \frac{1}{2}$ the rate $\Gamma$ is peaked at $E=0$, while
in the case ($z' > \frac{1}{2}$) the rate is zero at $E=0$ and
the maximum is reached at finite energy $\widetilde{E} \leq
4E_c$, or, in other terms, at finite positive voltage $V$ across
the junction. This property of $\Gamma(E)$ in circuits with
substantial dissipation plays the crucial role in the operation
of the three-junction R-pump.

For one junction in the three-junction array with gate
capacitances $C_g \ll C$ the effective impedance Re$Z(0) \equiv
R'=\frac{1}{9}R$, i.e. the parameter $z' = \frac{1}{9}z$. The
factor $\frac{1}{9}$, which considerably attenuates the effect of
resistors, stems from the square of the ratio of the total
capacitance of the series array ($=\frac{1}{3}C$) to the junction
capacitance $C$ (see a similar network analysis in Ref.
\cite{Ingold}).

Accordingly, the equivalent damping for two junctions of the
three-junction array is larger and determined by $R' =
\frac{4}{9}R$ and, hence, $z' =\frac{4}{9}z$. Therefore, the rate
of Cooper pair transition simultaneously across two junctions,
i.e. the rate of the co-tunneling, at small $E$ is
\begin{equation} \Gamma_{\rm cot}(E) \propto
\lambda^4 E^{\frac{8}{9}z-3}.
\end{equation}
This expression has been obtained by considering the environmental
effect in a similar fashion as was done by Odintsov $et~al.$
\cite{Odintsov} for the co-tunneling of normal electrons in a
single-electron transistor. Accordingly, co-tunneling across all
three junctions  decays even stronger ($\propto \lambda^6
E^{2z-5}$), because the whole array experiences the full
resistance $R$.

The comparison of Eq.~(6) with Eq.~(5) taken for the value
$z'=\frac{1}{9}z$ shows that the rate of co-tunneling is
drastically depressed if $z$ is sufficiently large. For example,
for $z=9$ (or $R \approx 58.5$~k$\Omega$) the rate of Cooper pair
co-tunneling across two junctions $\Gamma_{\rm cot} \propto E^5$
(i.e. it is similarly small as single-electron co-tunneling
across a three-junction normal array \cite{Averin89}). On the
other hand, the direct tunneling rate $\Gamma \propto E$, is
similar to that of ordinary electron tunneling. The effective
junction resistance in the latter case is $R_{\rm eff}\approx
0.03 \lambda^{-2} R_Q$. For the practical value $\lambda \sim
0.03$ this formula yields $R_{\rm eff}\sim 200$~k$\Omega$, i.e.
the typical value of a normal single electron junction. Since the
junction resistance determines the characteristic time constant
$RC$, it can be expected that the frequency characteristics of the
Cooper pair R-pump are at least not worse than those of the normal
electron counterpart without resistors.

\section{Experiment}

The Al/AlO$_x$/Al tunnel junctions and Cr resistors were
fabricated $in~situ$ by the three-angle shadow evaporation
technique through a trilayer mask patterned by e-beam lithography
and reactive-ion etching. The tunnel junction parameters of the
best suited available sample were found to be: capacitance
$C\approx 250$~aF and normal-state resistance
$R_j\approx160$~k$\Omega$. The former value yields $E_c\approx
320~\mu$eV. On the assumption that the Ambegaokar-Baratoff
relation between the critical current $I_c$ and the junction
resistance is valid, the aforementioned value of $R_j$ yields the
value $E_J = \Delta_{\rm Al}R_Q/2R_j\approx 4.1~\mu$eV, where
$\Delta_{\rm Al} \approx 200~\mu$eV is the superconducting energy
gap of aluminum at low temperature. These parameters of the
sample give the ratio of characteristic energies  $\lambda
\approx 0.013$.

The thin-film Cr resistors (thickness $d \approx 7$~nm) had
lateral dimensions of about 80~nm by 10~$\mu$m. For the purpose of
an independent characterization of the resistors and junctions, we
attached four resistors in twos to either end of the junction
array, resulting in the equivalent resistance $\frac{1}{2}R$ on
each side as depicted in Fig.~\ref{schema}. (For the measurement
of the I-V curves we used, however, the two-wire configuration.)
Either of four resistors had $R \approx 60$~k$\Omega$ and
reasonably low self-capacitance per unit length, $\approx
60$~aF/$\mu$m. The resistors exhibited negligibly small ($< 2$~\%)
non-linearity of the I-V curve in the region of small voltages at
the millikelvin temperature \cite{Zorin}.

\begin{figure}[h] \begin{center}
\includegraphics[width = .88\columnwidth]{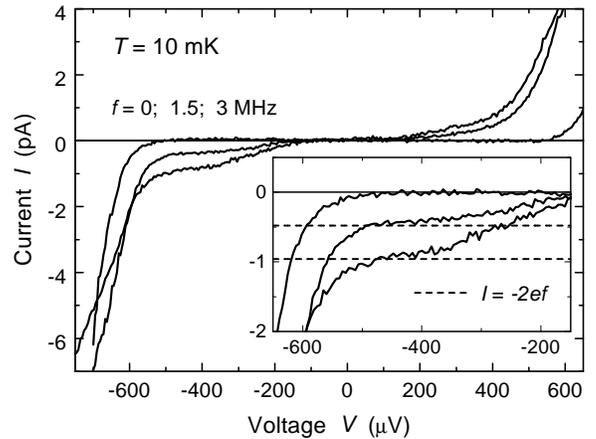}
\caption{I-V characteristics of the R-pump without and with ac
drive of two different frequencies. The inset shows the blow-up
of the steps whose positions were expected to be strictly at
$I=-2ef$ (shown by dashed lines), where the sign "-" is due to
the phase relation chosen between two gate signals.} \label{IVC}
\end{center}
\end{figure}

The samples were measured in the dilution refrigerator at the
temperature $T$ of about 10~mK, i.e. well below the critical
temperature of aluminum, $T_c\approx$ 1.15~K. The bias and gate
lines were equipped with pieces of the thermocoax cable which was
thermally anchored at the sample holder plate and served as a
filter for frequencies above 1~GHz \cite{thermocoax}. The typical
I-V curves with and without ac drive are presented in
Fig.~\ref{IVC}. The curves exhibit considerably smeared steps
whose position ($I\approx -2ef$) clearly shows the linear
current-versus-frequency dependence (although the quality of the
steps is noticeably degraded with increasing frequency)
\cite{Lotkhov}. Another feature of the steps is their position on
the voltage axes: they appear at finite voltage $V$ applied in
the current direction, while at $V=0$ neither co-tunneling nor
pumping effect were observed. We attribute this effect to the
damping effect of the resistors, which was too strong in this
sample. Thus, the behavior of the Cooper pair R-pump differs
significantly from that of the Cooper pair pump without resistors
\cite{Geerligs}.

\section{Discussion}
Although our first experiment evidenced the desirable effect of
Cooper pair pumping, a further improvement is inevitably needed.
First, the sample parameters should be optimized: Josephson
coupling should be increased (possibly up to $\lambda =0.1$)
while the resistance value $R$ should be somewhat reduced. These
parameter values lead to an increase of the tunneling rate and, as
a consequence, to Cooper pair pumping at zero voltage bias $V$.
Secondly, the quasiparticle tunneling, resulting in sporadic
translations of the cycle trajectory and, hence, to the pumping
errors, should be reliably depressed: As the measurements showed,
all our Al-Cr samples (as well as those with $E_c < \Delta_{\rm
Al}$) so far suffered from quasiparticle tunneling
\cite{e-periodicity} and did not show the so-called parity effect
\cite{Averin92}, \cite{Tuominen}. It is particularly remarkable
that our Cr resistors do not prevent an uncontrolled poisoning of
Josephson tunneling by non-equilibrium quasiparticles arriving at
the islands from the external circuit. Such "buffer" effect of
the normal electrodes was proposed and demonstrated in Al-Cu
devices by Joyez $et~al.$ \cite{Joyez}. That is why the problem
of the poisoning quasiparticle tunneling requires a more radical
solution. Probably manufacture of small-capacitance niobium
junctions ($\Delta_{\rm Nb} \approx 1.4$~meV $\gg E_c$) could
improve the situation by making the parity effect in these
structures stable.

In conclusion, we proposed a simple superconductor-normal metal
circuit enabling, in principle, the efficient pumping of Cooper
pairs. Due to the energy dissipation in the resistors, the effect
of Cooper pair co-tunneling is heavily damped. A possibility of
pair pumping was demonstrated and improvements of the experiment
are proposed.


\begin{acknowledgments}
This work is supported in part by the EU Project COUNT.
\end{acknowledgments}


\begin{references}
\bibitem{e-periodicity}
This was found from the current-versus-gate voltage dependencies
whose behavior reflected the charging of the islands with the
charge quantum $q=e$.
\end{references}
\end{document}